# A two-phase approach for detecting recombination in nucleotide sequences

CHEONG XIN CHAN
The University of Queensland

ROBERT G. BEIKO
Dalhousie University
and

MARK A. RAGAN
The University of Queensland

___

Genetic recombination can produce heterogeneous phylogenetic histories within a set of homologous genes. Delineating recombination events is important in the study of molecular evolution, as inference of such events provides a clearer picture of the phylogenetic relationships among different gene sequences or genomes. Nevertheless, detecting recombination events can be a daunting task, as the performance of different recombination-detecting approaches can vary, depending on evolutionary events that take place after recombination. We recently evaluated the effects of post-recombination events on the prediction accuracy of recombination-detecting approaches using simulated nucleotide sequence data. The main conclusion, supported by other studies, is that one should not depend on a single method when searching for recombination events. In this paper, we introduce a two-phase strategy, applying three statistical measures to detect the occurrence of recombination events, and a Bayesian phylogenetic approach in delineating breakpoints of such events in nucleotide sequences. We evaluate the performance of these approaches using simulated data, and demonstrate the applicability of this strategy to empirical data. The two-phase strategy proves to be time-efficient when applied to large datasets, and yields high-confidence results.

___

## 1. INTRODUCTION

Genetic recombination is the process in which an external fragment of a genetic material is integrated into a recipient sequence. In eukaryotes, recombination plays a critical role in ensuring proper pairing and correct segregation of chromosomes during meiosis, and is important in maintaining genome integrity throughout cell division. Intuitively, recombination depicts a molecular process in which regions of a continuous piece of DNA are disassociated or shuffled. The process of recombination, or genetic transfer in general, contributes to genetic diversity and inconsistent phylogenetic signals across genomes of different species. Therefore, elucidating genetic transfers resulting from recombination events in biological sequences will enhance our understanding of the role selective forces play in shaping genomes. However, detecting recombination is not without problems. When the recombining sequences are very similar to each other, or when subsequent evolution has obscured the recombination signal, detecting recombination can be difficult. The scenario is more complicated when there are overlapping recombination events on the sequences within proximity of each other.

  A number of approaches are available for detecting recombination events in biological sequences. These approaches can be classified based on the different algorithms adopted: distance-based [Weiller 1998; Etherington et al. 2005], substitution distribution-based [Sawyer 1989], compatibility-based [Jakobsen and Easteal 1996; Bruen et al. 2006], and phylogenetic-based [Hein 1990]. New approaches are also being developed such as one adopting genetic algorithm [Pond et al. 2006] and one combining the use of two different statistical tests [Graham et al. 2005]. A number of these approaches were reviewed for their performance; the effects of sequence divergence, amount of recombination, and subsequent substitutions after the recombination event were examined [Posada and Crandall 2001; Wiuf et al. 2001; Posada 2002; Chan et al. 2006]. All these studies agree that recombination is easier to detect when the event involves

___

sequences that are divergent, and that approaches based on compatibility and substitution distribution showed higher prediction accuracy than the conventional phylogenetic approach. The importance of not depending on a single method in isolation when detecting recombination was well demonstrated in these studies, as the different approaches have different advantages and drawbacks. A good method in detecting occurrence of recombination might not be good in identifying the breakpoints of such events, and vice versa. Using different approaches in succession can increase our confidence in detecting an event that has indeed taken place within the defined breakpoints at the sequences. When dealing with large datasets, it is desirable to use a quick method in first-pass screening for the presence of recombination in the datasets, then a more-accurate (albeit slower) method to delineate the recombination breakpoints among the positives [Chan et al. 2006]. Here we present a two-phase strategy in detecting recombination events in nucleotide sequences, and highlight its application in detecting large-scale recombination.

## 2. TWO-PHASE APPROACH FOR DETECTING RECOMBINATION

The strategy involves two phases: (a) detecting occurrence of recombination events in the sequence dataset; and (b) identifying breakpoints of such events. The first phase involves a quick, first-pass screening for possible recombination events using a number of statistical measures for evaluating phylogenetic discrepancy across the set of sequences. Once significant phylogenetic discrepancy is detected, the second phase involves a slower but more-accurate Bayesian phylogenetic approach to delineate recombination breakpoints in sequence data.

### 2.1 Phase I: Detecting occurrence of recombination

For first-pass screening for occurrence of recombination events, we implemented three different statistical measures: neighbour similarity score (NSS) [Jakobsen and Easteal 1996], maximal chi-squared (MaxChi) [Maynard Smith 1992] and pairwise homoplasy index (PHI) [Bruen et al. 2006]. All three were implemented in PhiPack [Bruen et al. 2006]. These statistics measure the significance of phylogenetic discrepancy across sites in an alignment, each test with an assigned p-value. Both NSS and PHI are based on compatibility of parsimoniously informative sites, whereas MaxChi is based on substitution distributions across sites. If all three p-values show high significance (each p-value < 0.05), recombination is most likely present within the sequence set. While we use these three measures, one could imagine using other substitution distribution-based methods during this phase.

### 2.2 Phase II: Identification of recombination breakpoints

After the sequence set is found to be positive in first-pass recombination screening, recombination breakpoints can then be identified with high accuracy using other approaches, e.g. Bayesian phylogenetic approach [Chan et al. 2006]. We used DualBrothers [Minin et al. 2005], which implements a Bayesian approach using reversible jump Markov chain Monte Carlo and dual multiple change-point model in inferring changes in tree topologies and evolutionary rates across sites within a sequence set. While prediction accuracy of recombination breakpoints comes at the expense of time and computational resources, the two-phase strategy avoids the use of time-consuming approaches in delineating recombination on sequence datasets that are potentially negatives for recombination in the first place.

## 3. EVALUATION OF PERFORMANCE

To evaluate the performance of the approaches used in our two-phase strategy, we simulated data with a single recombination event in the middle of a four-sequence set. The effects of subsequent substitutions after recombination and the sequence divergence prior to recombination were assessed. We used Seq-Gen [Rambaut and Grassly 1997] to simulate sequence evolution. Four-taxon sequence sets of length 1000 nt were generated using the HKY model of substitution [Hasegawa et al. 1985] with nucleotide frequencies A = 0.20, C = 0.30, G = 0.30, T = 0.20, a transition/transversion ratio of 2, and a four-category discrete approximation to a gamma distribution of among-site rate variation with shape parameter alpha = 1.0. Different evolutionary histories prior to recombination, and different amount of subsequent substitutions after the recombination event were used in the simulations, and a total of 100 replicates were used for each combination. The recombination event was simulated by exchanging or replacing the fragments within the sequences resembling a reciprocal or non-reciprocal event respectively. After recombination, each lineage was evolved independently of the other with different amount of subsequent substitutions per site. See [Chan et al. 2006] for more details on how the simulated sequence sets were generated.

The prediction accuracy for each of the three statistical measures implemented in PhiPack was evaluated by the p-value generated; a small p-value implies that the phylogeny discrepancy within the sequence set is significant, hence recombination is highly probable. Figure 1 shows the proportion of positive simulation sets for recombination that were assigned significant p-values ($\leq 0.05$); across different combinations of prior evolutionary histories (showing sequence divergence) and subsequent substitutions. The notation 'L05/50' represents internal branch length of 0.05 and external branch length of 0.50 on the simulated tree topology before the recombination event, the units being number of substitutions per site. Longer external branch lengths depict that the sequences are more divergent prior to recombination.

As shown in Figure 1, as more subsequent substitutions were simulated, the proportion of simulated sets showing significant p-values decreased accordingly (e.g. NSS, proportion < 0.4 when subsequent substitution is 0.5 in L50/05). In general, MaxChi and PHI showed higher accuracy compared to NSS. Significant recombination events are detected in over 80% of the simulated set using MaxChi, even when subsequent substitutions exceeded 0.25 substitutions per site. However, in situations where recombination involves closely related sequences e.g. in cases of L05/05, the recombination events are more difficult to recover, particularly when the amount of subsequent substitutions exceeded 0.25; the percentage of significant detection in all three measures was < 40%. This finding supports previous studies which found that recombination event involving closely related sequences are more difficult to detect than similar events involving sequences which are more divergent.

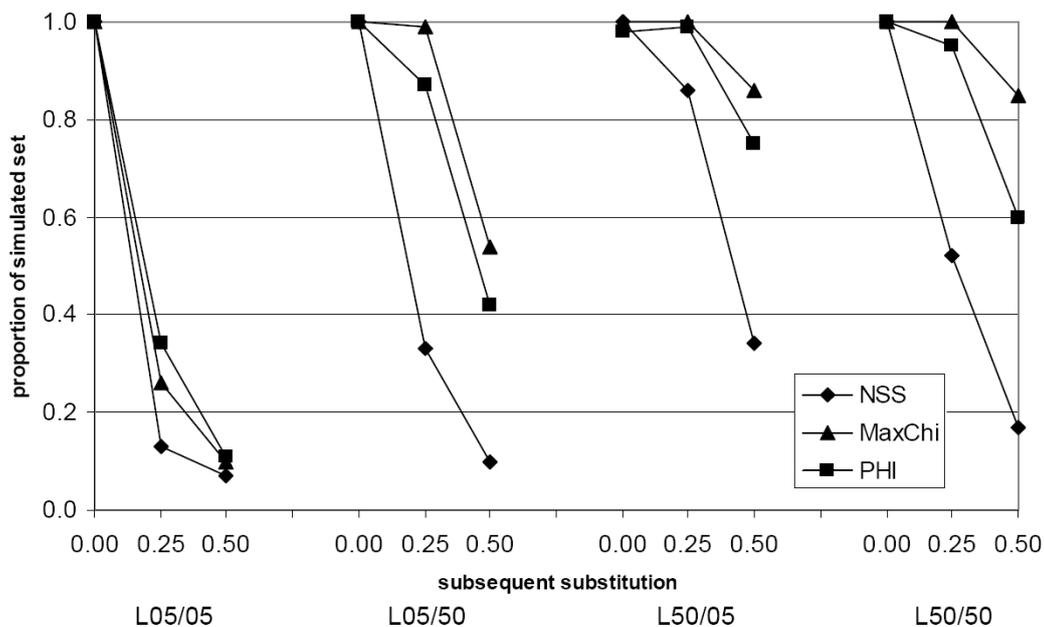

*Figure 1. Prediction accuracy of PhiPack on simulated data. The Y-axis shows the proportion of simulation set that was assigned a p-value of ≤0.05, for each of the three statistical measures: neighbour similarity score (NSS), maximal chi-squared (MaxChi) and pairwise homoplasy index (PHI), across different prior evolutionary histories and substitutions on the sequences after recombination. Results shown for non-reciprocal recombination events.*

Although the prediction accuracy of these statistical measures was affected by the effects of subsequent substitutions and prior evolutionary history (sequence divergence), PHI was found to be less sensitive to the amount of subsequent substitutions and the prior evolutionary histories of the sequence sets (multiple regression analysis, adjusted $R^2$ 0.341, F statistic 156.1, 1195 degrees of freedom) compared to the other approaches in our previous study [Chan et al. 2006] e.g. adjusted $R^2$ 0.85 for GENECONV, a substitution distribution-based method. All three statistical measures are quick and less dependent on parameter settings compared to other approaches e.g. GENECONV (substitution distribution-based) or RecPars (phylogenetic-based). The measures also showed low false positive discovery rates on the control sequence sets that were void of recombination event (results not shown). By taking three different statistical measures into account, recombination events can be positively detected at a higher level of confidence than is the case when a single method is used in isolation. PhiPack proved to be an efficient tool to perform this task.

For identifying recombination breakpoints, we used DualBrothers, a Bayesian phylogenetic-based approach using a multiple change-point model described by eight parameters related to location of breakpoints, tree topologies and evolutionary rates. Using the same simulated dataset, the algorithm was found to be accurate in delineating recombination breakpoints, but at the expense of computational resources and time [Chan et al. 2006]. Therefore, having a first-pass screening in the recombination-detecting strategy will save time and computational resources, such that the accurate-but-slow method needs only be implemented on those sequence sets that show evidence of recombination.

## 4. APPLICATION TO EMPIRICAL DATA

We applied the two-phase strategy in an attempt to infer recombination events among families of protein-coding sequences among prokaryotic genomes. The dataset consisted of 22437 putatively orthologous protein families obtained from 144 fully sequenced prokaryotic genomes [Beiko et al. 2005]. Clustered via a hybrid approach of naïve and Markov clustering algorithms [Harlow et al. 2004], protein alignments of these gene families were validated using a pattern-centric objective function [Beiko et al. 2005], and converted into nucleotide sequence alignments for the analysis of recombination.

The first-pass screening for occurrence of recombination within the DNA alignments was carried out using the three statistical measures in PhiPack, in which detection is treated as positive when two out of three measures (NSS, MaxChi and PHI) yield a p-value ≤ 0.10. Out of 1462 DNA alignments of strictly orthologous gene families, 427 (29.2%) showed evidence of recombination, fulfilling this criterion. The quick screening step greatly reduced the number of alignments needed for breakpoint detection in the next step of the strategy. For identification of breakpoints, DualBrothers was implemented on each the 427 DNA alignments with MCMC chain length = 1020000 generations, burn-in at 20000 generations, and window length = 5.

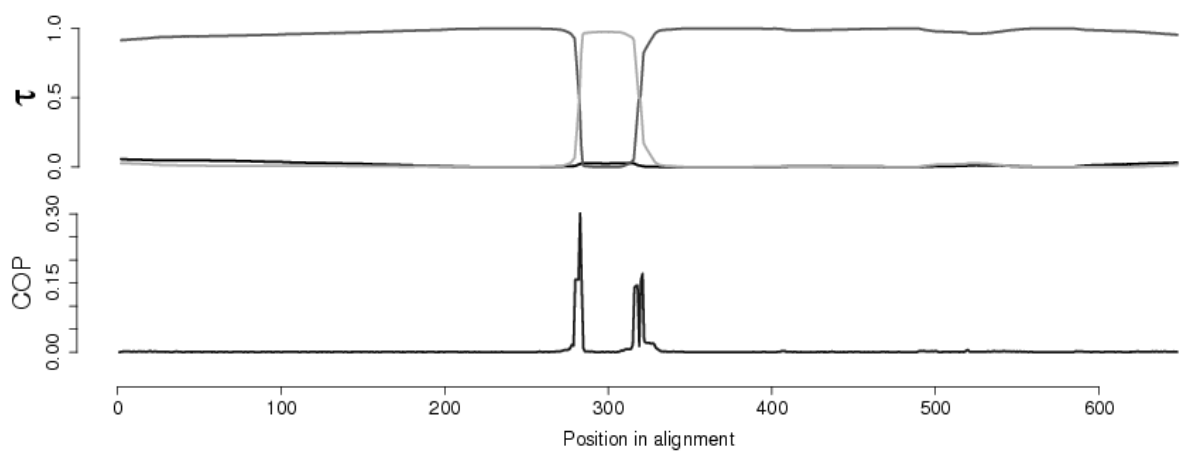

*Figure 2. Identification of recombination breakpoints in gene family s1605. The upper graph shows posterior probability of different tree topology across the sites in the alignment; each line represents a different tree topology. The lower graph shows the COP (change-of-point) marginal posterior probability across sites in the alignments. A site with a sharp peak is probably a recombination breakpoint.*

Figure 2 shows an example of a recombination event detected in a gene family, s1065, a family of hypothetical proteins consisting four sequences. As shown by the peaks in the change-of-point (COP) marginal posterior probability (mPP) plot and the corresponding change of tree topology, two possible breakpoint positions were detected, one between alignment positions 250-300, and another between positions 300-350. The exact breakpoints were determined by a statistical approach involving sub-sampling of COP mPP under a peak with 95% Bayesian confidence interval [Suchard et al. 2003; Minin et al. 2005], as shown in Table 1. Positions 282 and 319 in the DNA alignments were found to be where the recombination breakpoints are located, separating the sequences into three partitions. Figure 3 shows the dominant tree topology within each partition of the alignment.

| Position in alignment | |  |
|---|---|---|
| COP (Breakpoint) | 95% Bayesian Confidence Interval | |
| 282 | 277 | 284 |
| 319 | 312 | 329 |

*Table 1. Inferred breakpoints in gene family s1605.*

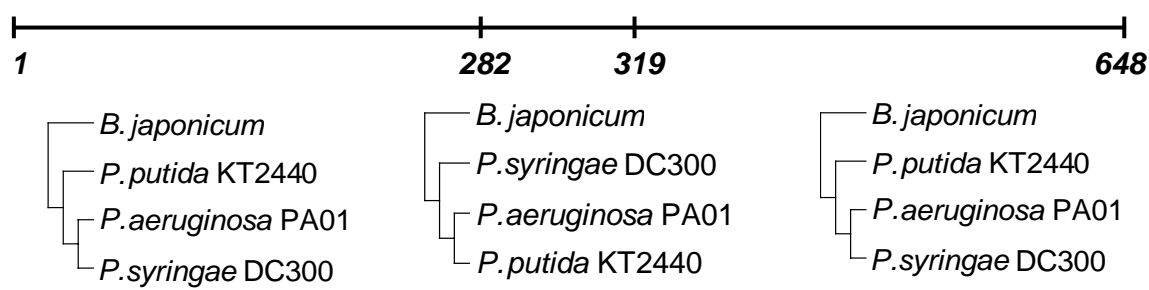

*Figure 3. Tree topology within each partition separated by the recombination breakpoints. The line on top of the tree topologies depicts the positions of the DNA alignment.*

As shown in Figure 3, a recombination was detected in this gene family, in which the recombined region was inferred in the middle of the alignment between positions 282-319. By examining the tree topologies, a possible reciprocal recombination event can be inferred between lineage *Pseudomonas syringae* DC300 and lineage *Pseudomonas putida* KT2440.

## 5. CONCLUSIONS

Using different approaches in succession, we were able to detect recombination events with higher confidence than if a single method had been used in isolation. The first step of first-pass screening is quick and useful for filtering out sequence sets that show no evidence of recombination, making this approach suitable for detecting recombination in multi-genome scale data.

## 6. ACKNOWLEDGMENTS

This study is supported by an Australian Research Council (ARC) grant CE0348221. We thank Aaron Darling and Vladimir Minin for invaluable input on the use of DualBrothers and for providing a number of software codes. CXC is supported by a UQIPRS scholarship for his postgraduate study.